\documentclass[twocolumn,showpacs,preprintnumbers,amsmath,amssymb]{revtex4}
\usepackage{amsmath,epsfig,subfigure}
\usepackage{graphicx}
\usepackage{dcolumn}
\usepackage{bm}

\newcount\EquaNo \EquaNo=0
\def\newEq#1{\advance\EquaNo by 1 #1=\EquaNo}
\newcount\TablNo \TablNo=0
\def\newTabl#1{\advance\TablNo by 1 #1=\TablNo}
\newcount\FigNo \FigNo=0
\def\newFig#1{\advance\FigNo by 1 #1=\FigNo}
\newcount\ChapNo \ChapNo=0
\def\newCh#1{\advance\ChapNo by 1 #1=\ChapNo}
\def\expec#1{\big\langle{#1}\big\rangle}
\begin {document}

\title { Infinitely-fast diffusion in {\it {Single-File Systems}}  }

\author{S.V. Nedea}
\email{silvia@win.tue.nl}
\altaffiliation{
Department of Mathematics and Computer Science, Technical University of Eindhoven
}

\author{A.P.J. Jansen}
\altaffiliation{ 
 Department of Chemical Engineering, Technical University of Eindhoven
}

\author{J.J. Lukkien}
\altaffiliation{
Department of Mathematics and Computer Science, Technical University of Eindhoven
}


\date{\today}

\begin {abstract}

  We have used Dynamic Monte Carlo(DMC) methods and analytical techniques to
analyze Single-File Systems for which diffusion is infinitely-fast. We have
simplified the Master Equation removing the fast reactions and we have
introduced a DMC algorithm for infinitely-fast diffusion. The DMC method for
fast diffusion give similar results as the standard DMC with high diffusion rates. 
 We have investigated the influence of characteristic parameters, such as pipe
length, adsorption, desorption and conversion rate constants on the
steady-state properties of Single-File Systems with a reaction, looking at
cases when all the sites are reactive and when only some of them are reactive.
 We find that the effect of fast diffusion on single-file
properties of the system is absent even when diffusion is infinitely-fast.
 Diffusion is not important in these systems.
 Smaller systems are less reactive and the occupancy profiles for
infinitely-long systems show an exponential behavior.
 
\end{abstract}
\pacs{02.70.Uu, 02.60.-x, 05.50.+q, 07.05.Tp}
\maketitle

\section { Introduction } 

    In one-dimensional systems such as zeolites or other porous structures,
diffusion is a very important process. The pores of these structures that
have the cross section somewhat larger than a benzene molecule, are modelled by
{\it {Single-File Systems}}. In these systems particles move in a concerted 
fashion, as they are unable to cross each other. This process of
{\it {Single-File}} diffusion has different characteristics from ordinary diffusion which
affects the nature of both transport and conversion by chemical reactions.
 In~\cite{silvia} and ~\cite{silvia1} we have studied the steady-state and transient properties of
this system. We have analysed different situations for diffusion rates and we
have compared the results obtained from simulation and analytical techniques.
 Often diffusion is a very fast process compared to the other reactions in the
system. We are thus interested to be able to model correctly infinitely-fast diffusion.
 For this purpose, we used DMC methods with high regular diffusion rates,
assuming that these rates are high enough to model infinitely-fast diffusion.
 
 Dynamic Monte Carlo methods for very high rates are not very efficient and
the progress of the simulation is slow.
 Moreover, considering regular reactions rates it is always a problem to
balance between making the diffusion rates high enough so that the infinitely-fast diffusion effects are 
correctly modelled and the performance of the simulation.
 We derive here a new method to simulate infinitely fast-diffusion in
{\it {Single-File Systems}}, starting from the Master Equation.

  The rate equations of some special limiting cases 
and an analytical description for the productivity of the system are also
derived.
 We study also how the system behavior changes for different sets of kinetic
parameters and different distributions of the reactive sites.
 We categorize also interesting results obtained for profile occupancies for 
different reactive system and different distribution of the reactive sites.

 In section ~\ref{sec:lev0} we specify our mathematical model with the
theoretical background for the analytical and simulation results.
 We introduce the Master Equation of the systems in section ~\ref{sec:lev2} and then we
simplify the Master Equation removing the fast reactions in section~\ref{sec:lev3}.
 In section~\ref{sec:lev4} and ~\ref{sec:lev5} we present the simulation methods and we present a DMC algorithm for
infinitely-fast diffusion. Different analytical results are presented 
in ~\ref{sec:lev6}. In section ~\ref{sec:lev10} and ~\ref{sec:lev11} we analyze different simulation results for the case
when all the sites are reactive and when only some of the sites are
reactive.
 We pay special attention to the influence of the length of the pipe and
reaction rate constant on the site occupancy of the system.

\section {\label{sec:lev0} Theory }

   In this section we will give the theoretical background for our analytical
and simulation results. First we will specify our model and we will
derive a finite set of exact rate equations starting from the Master
Equation~\cite{kampen}. These rate equations are used 
in order to derive expressions for the productivity in the system for special
cases. We show that we can simplify the Master Equation describing the
evolution of the system over time removing fast reactions.
 We use Dynamic Monte Carlo method for our simulation results and we
give the description of a Dynamic Monte Carlo-like algorithm for infinitely-fast
diffusion.

\subsection {\label{sec:lev1}The Model}

    We model a {\it Single-File System} by a one-dimensional array of
sites, each possibly occupied by an adsorbate. This is the model of
diffusion and reaction in a one-dimensional arrangements of particles with
hard-core interaction. The sites are numbered $1,2,\ldots ,S$. A particle can
 only move to the right or to the left if an adjacent site is vacant. 
 The sites could be reactive and unreactive and we note
with $N_{\rm prot}$ the number of reactive sites. A reactive site is the only place
where a conversion may take place.

\begin{figure*}
\centering
\subfigure {\epsfig {figure=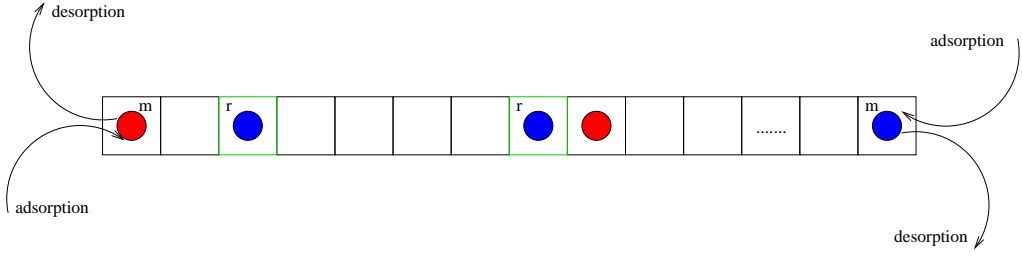, width=13.5cm} }
\caption {Picture of a {\it Single-File System} with two types of adsorbed particles}
\end {figure*}


       We consider two types of adsorbates, $\rm A$ and $\rm B$, in our model and we denote
with $\rm Y$ the site occupation of a site, $\rm Y$=($*$, $\rm A$, $\rm B$), which stands for an
vacant site, a site occupied by $\rm A$, or a site occupied by a $\rm B$, respectively.
   We restrict ourselves to the following mono and bi-molecular transitions.
\\
\\
a) Adsorption and desorption
\\
\\
    Adsorption and desorption take place only at the two marginal sites
i.e., the left and rightmost sites at the ends of the system.
\\
\begin{center}
    ${\rm A}({\rm gas})$ + $\rm *_{\it m}$ $\longrightarrow$ ${\rm A}_{\it m}$
\\
    ${\rm A}_{\it m}$   $\longrightarrow$   ${\rm A}({\rm gas})$ + ${\rm *}_{\it m}$
\\
    ${\rm B}_{\it m}$ $\longrightarrow$ ${\rm B}({\rm gas})$ + $*_{\it m},$
\\
\end{center}
where subscripts $\it m$ denotes a marginal site. Note that there is no $\rm B$ adsorption.
$\rm B$'s can only be formed  by a conversion.
\\
\\
b) Diffusion
\\
\\
    In the pipe, particles are allowed to diffuse via hopping to
    vacant nearest neighbor sites.
\\
\begin{center}
   ${\rm A}_{\it n}$ + ${\rm *}_{\it n+1}$ $\longleftrightarrow$ ${\rm *}_{\it n}$ + ${\rm A}_{\it n+1}$
\\
   ${\rm B}_{\it n}$ + ${\rm *}_{\it n+1}$ $\longleftrightarrow$ ${\rm *}_{\it n}$ + ${\rm B}_{\it n+1}$
\\
\end{center}
where the subscripts are site indices: $n$=$\textsl{1, 2, \ldots, S-1}$.
\\
\\
c) Conversion
\\
\\
   An $\rm A$ can transform into a $\rm B$ at a reactive site.
\\
\begin {center}
    $\rm A_{\it r}$ $\longrightarrow$ $\rm B_{\it r}$.
\end{center}

       In the initial state of the system all the sites are vacant (no particles 
in the pipe) as we are interested in the behavior of the system towards equilibrium.

\subsection {\label{sec:lev2} Master Equation}

  Reaction kinetics is described by a stochastic process. Every reaction has a microscopic
rate constant associated with it that is the probability per unit time that
the reaction occurs.
  Stochastic models of physical systems can be described by a Master
Equation.~\cite{kampen}

 By $\alpha$, $\beta$, we will indicate a particular configuration of the
system i.e., a particular way to distribute adsorbates over all the sites.
 $P_\alpha(t)$ will indicate the probability of finding the system in
configuration $\alpha$ at time $t$ and $W_{\alpha\beta}$ is the rate
constant of the reaction changing configuration $\beta$ to configuration
$\alpha$.

    The probability of the system being in configuration $\alpha$ at time
$t+dt$ can be expressed  as the sum of two terms. The first term is the
probability to find the system already in configuration $\alpha$ at time $t$
multiplied by the probability to stay in this configuration during $dt$.
 The second term is the probability to find the system in some other
configuration $\beta$ at time $t$ multiplied by the probability to go from
$\beta$ to $\alpha$ during $dt$.

\begin{equation}
P_{\alpha}(t+dt)=(1-dt\sum_{\beta} W_{\beta\alpha})P_{\alpha}(t) +
         dt\sum_{\beta}W_{\alpha\beta}P_{\beta}(t)
\end {equation}

By taking the limit $dt \to 0$ this equation reduces to a Master Equation:

\begin{equation}
  {dP_\alpha(t)\over dt}
  =\sum_{\beta}
   \left[W_{\alpha\beta}P_{\beta}(t)-W_{\beta\alpha}P_\alpha(t)\right].
\end{equation}

Analytical results can be derived as follow. The value of a property $X$ is a weighted average over the values
$X_{\alpha}$ which is the value of $X$ in configuration $\alpha$:

\begin{equation}
   \langle X \rangle=\sum_{\alpha}P_{\alpha}X_{\alpha}.
\end{equation}

  From this follows the rate equation
\begin{equation}
\begin{split}
 {{d\langle X \rangle}\over{dt}}
  & =\sum_{\alpha}{dP_{\alpha}\over{dt}}X_{\alpha}\cr
  &=\sum_{\alpha\beta}[W_{\alpha\beta}P_{\beta}-W_{\beta\alpha}P_{\alpha}]X_{\alpha}\cr
  &=\sum_{\alpha\beta}W_{\alpha\beta}P_{\beta}(X_{\alpha}-X_{\beta}).
\end{split}
\end{equation}

\subsection{\label{sec:lev3}Master Equation for infinitely-fast diffusion}

 We show that we can simplify the Master Equation removing the fast
reactions.
 In order to remove fast reactions we stop distinguishing between
configurations that can be transformed into each other by the fast
reactions. We split all configurations into disjoint sets such that if
$C_i$ is one such a set and $\alpha, \beta \in C_i$, then $\alpha$ can be
transformed into $\beta$, or vice versa by fast reactions.
 If we denote $${\pi}_i=\sum_{\alpha \in C_i}P_{\alpha},$$ we can derive the 
Master Equation for $\pi_i$.

\begin{equation}
\begin{split}
{{d\pi_i}\over{dt}}&=\sum_{\alpha \in C_i}{dP_{\alpha}\over{dt}}\cr
 &= \sum_{\alpha \in C_i}\sum_{\beta}{\left[ W_{\alpha\beta}
P_{\beta}-W_{\beta\alpha}P_{\alpha}\right]}\cr
 &=\sum_{\alpha \in C_i}\sum_{j}\sum_{\beta \in C_j}\left[W_{\alpha\beta}P_{\beta}-W_{\beta\alpha}P_{\alpha}\right]\cr
 &=\sum_{j} \left(\sum_{\alpha \in C_i}\sum_{\beta \in C_j}W_{\alpha\beta}{{P_{\beta}}\over{\pi_j}}\right)\pi_j \cr
 & - \sum_{j} \left( \sum_{\beta \in C_j}\sum_{\alpha \in C_i}W_{\beta\alpha}{{P_{\alpha}}\over{\pi_i}}\right)\pi_i \cr
 &=\sum_j\left[\omega_{ij}\pi_j -\omega_{ji}\pi_i\right]
\end{split}
\end{equation}

with $\omega_{ij}=\sum_{\alpha \in C_i}\sum_{\beta \in C_j}{W_{\alpha\beta}
{P_{\beta}\over{\pi_j}}}.$

  We see that all fast reactions have disappeared; they only contribute to 
$\omega_{ii}$, which can be left out of the ME.
  In order to calculate $\omega_{ij}$, we need the conditional probabilities
$P_{\beta}\over{\pi_j}$. Because we have fast reactions connecting
the $\beta$'s in $C_j$ we may assume that these $\beta$'s are in
steady-state with respect to each other. Hence, the conditional probability
$P_{\beta}\over{\pi_j}$ is nothing but the probability of $\beta$ in
steady-state if we restrict ourselves to the configurations in $C_j$.

\subsection {Simulation methods}

\subsubsection {\label{sec:lev4}Dynamic Monte Carlo}

 DMC methods allow us to simulate the system governed by the
Master Equation over time.
  We simplify the notation of the Master Equation by defining a matrix $\bf W$
containing the rate constants $W_{\alpha\beta}$, and a diagonal matrix $\bf R$ by
${R}_{{\alpha}{\beta}}\equiv \sum_{\gamma}W_{\gamma\beta}$, if ${\alpha}={\beta}$,
and 0 otherwise.
 If we put the probabilities of the configurations $P_{\alpha}$ in a vector
$\bf P$, we can write the Master Equation as

\begin {equation}
 {d{\bf P}\over{dt}}=-({\bf R}-{\bf W}){\bf P}.
\end{equation}
where $\bf R$ and $\bf W$ are time independent.
We also introduce a new matrix $\bf Q$, ${\bf Q}(t) \equiv \exp[-{\bf R}t].$

This matrix is time dependent by definition and we can rewrite the Master
Equation in the integral form

\begin{equation}
{\bf P}(t)={\bf Q}(t){\bf P}(0)+\int_0^tdt^{\prime}{\bf Q}(t-t^{\prime}){\bf W}{\bf P}(t^{\prime}).
\end{equation}
By substitution we get from the right-hand-side  for $P(t^{\prime})$

\begin{equation}
\begin{split}
{\bf P}(t)
  & =[{\bf Q}(t)\cr
  & +
     \int_0^t dt^{\prime}{\bf Q}(t-t^{\prime}){\bf W}{\bf Q}(t^{\prime})\cr
  & +
    \int_0^tdt^{\prime}\int_0^{t^{\prime}}dt^{\prime\prime}{\bf Q}(t-t^{\prime}){\bf W}{\bf Q}(t^{\prime}-t^{\prime\prime})
    {\bf W}{\bf Q}(t^{\prime\prime})\cr
  & +\ldots]{\bf P}(0).
\end{split}
\end{equation}

 Suppose at $t=0$ the system is in configuration $\alpha$ with probability
$P_{\alpha}(0)$. The probability that, at time $t$, the system is still in
configuration $\alpha$ is given by
$Q_{\alpha\alpha}(t)P_{\alpha}(0)=\exp(-R_{\alpha\alpha}t)P_{\alpha}(0)$.
 This shows that the first term represents the contribution to the
probabilities when no reaction takes place up to time $t$. The matrix $\bf W$
determines how the probabilities change when a reaction takes place. The
second term represents the contribution  to the probabilities when no
reaction takes place between times $0$ and $t^{\prime}$, some reaction takes
place at time $t^{\prime}$, and then no reaction takes place between
$t^{\prime}$ and $t$. The subsequent terms represent contributions when two,
three, four, etc. reactions take place.
 The idea of the DMC method is not to compute probabilities
$P_{\alpha}(t)$ explicitly, but to start with some particular configuration,
representative for the initial state of the experiment one wants to
simulate, and then generate a sequence of other configurations with the
correct probability.
 The method generates a time $t^{\prime}$ when the first reaction occurs
according to the probability distribution $1-\exp[-R_{\alpha\alpha}t]$.
At time $t^{\prime}$ a reaction takes place such that a new configuration
${\alpha}^{\prime}$ is generated by picking it out of all possible new
configurations $\beta$ with a probability proportional to
$W_{{\alpha}^{\prime}{\alpha}}$. At this point we can proceed by repeating
the previous steps, drawing again a time for a new reaction and a new
configuration.~\cite{lukkien, gelten_all}
 One of the most popular DMC method in the literature is Random Selection
Method(RSM)~\cite{lukkien}. We use this method to simulate the Master Equation of our
system.

\subsubsection {\label{sec:lev5} A Dynamic Monte Carlo algorithm for infinitely-fast diffusion}

  In section C we have derived the ME distinguishing between configurations
that can be transformed into each other by fast reactions.
  Starting from the ME(5) we give a DMC algorithm for simulating
infinitely-fast diffusion.
  For our model, diffusion is much faster than adsorption and desorption, so
the sets are all configurations that are connected by diffusion.
  These sets can be labeled only by the sequences of particles ${\rm A}$ and
${\rm B}$, as only the number of $\rm A$ and $\rm B$ particles and their order in the
pipe is important to distinguish the configurations within a set.  Moreover, all
probabilities of configurations within a set are the same. This means that 
${\pi_j}\over{P_{\beta}}$ is the number of configurations in $C_j$. There are
$\begin{pmatrix}S\cr n \end{pmatrix}$ways to distribute $n$ particles over
$S$ sites. We have then 

$${P_{\beta}\over{\pi_n}}=\begin{pmatrix} S\cr n \end{pmatrix}^{-1}, $$
with $P_{\beta}\in C_n.$
 The summation $\sum_{\beta \in C_n }$ sums over $\begin{pmatrix}S\cr n \end{pmatrix}$
configurations. However, for adsorption the left-most or the right-most
site should be vacant.This gives us $\begin{pmatrix}S-1\cr n \end{pmatrix}$
possible configurations. Each of these gives just one configuration in the
summation over $\alpha$. So we get

\begin{equation}
\omega_{ads}=
2W_{ads}{\begin{pmatrix}S-1\cr n \end{pmatrix}\over{\begin{pmatrix}S\cr n \end{pmatrix}}}
={2W_{ads} {{S-n}\over{S}}}
=2W_{ads}(1-\theta),
\end{equation}
where $\omega_{ads}$ is the transition probability for the transition from a
configuration within the set with $n$ particles to a configuration within the
set with $n+1$ particles.
 Similary we find that 
\begin{equation}
\omega_{des}=
2W_{des}{\begin{pmatrix}S-1\cr n-1 \end{pmatrix}\over{\begin{pmatrix}S\cr n
\end{pmatrix}}}={2W_{des} {n}\over{S}}=2W_{des}\theta.
\end{equation}
In both expressions ${\theta}={{n}\over {S}}$ is the coverage.

The Dynamic Monte Carlo (DMC) algorithm that we have used to simulate the
system consists of the following steps:
\\
1. Compute the time for the next adsorption or desorption. If the
  current time is $t$ then the time for that process is $t+\Delta t$
  with
\begin{equation}
  \Delta t=-{\ln r\over 2(1-\theta)W_{\rm ads}+2\theta W_{\rm des}}
\end{equation}
where $r$ is a random number picked from the uniform distribution on the
interval $\langle 0,1]$ and $\theta$ is the probability that the
marginal site is occupied. With infinitely fast diffusion this
probability is given by $\theta=(N_{\rm A}+N_{\rm B})/S$ with $N_{\rm
A}$ and $N_{\rm B}$ the number of $\rm A$'s respectively $\rm B$'s in the system.
\\
2. Compute for each $\rm A$ in the system a time when it will transform
  into a $\rm B$. This time is given by $t+\tau$ with
\begin {equation}
  \tau=-{\ln r\over P W_{\rm rx}}.
\end{equation}
In this expression $P$ stands for the probability that the A is at a
protonic site. If we number the particles in the system from left to
right $1,2,\ldots,N_{\rm A}+N_{\rm B}$ and the sites also from left to
right $1,2,\ldots,S$ then the probability that particle number $n$ is at
site number $s$, $P_{n}^s$, is given by
\begin{equation}
 {P_{n}^s}=
  {\begin{pmatrix} s-1\cr n-1\cr \end{pmatrix}
   \begin{pmatrix} S-s\cr N_{\rm A}+N_{\rm B}-n\cr \end{pmatrix}
  \over\begin{pmatrix} S\cr N_{\rm A}+N_{\rm B}\cr \end{pmatrix}}.
\end{equation}
$P$ for particle $n$ is then the sum of this expression over all
protonic sites
\begin{equation}
P=\sum_{s=1}^S {P_{n}^s}{\delta}_s,
\end{equation}
where ${\delta}_s=1$ if site $s$ is protonic.
\\
3. Change those $\rm A$'s with $\tau<\Delta t$ into $\rm B$'s.
\\
4. Determine the next process at the marginal sites. It is an
  adsorption with probability proportional to $(1-\theta)W_{\rm ads}$
  and a desorption with probability proportional to $\theta W_{\rm
  des}$. The process is equally likely to occur on the left- or the
  right-hand-side.
\\
5. Change the number of particles in the system according to the
  next process at the marginal sites.
\\
6. Update the time.
\\
7. Repeat steps 1 to 6.

\section {\label{sec:lev6}Analytical results}

 In this section, for some special cases such as low loading limit
and fast and slow reaction, we are able to derive some expressions for the 
productivity in the steady-state. For the case of low loading limit we can
also derive the rate equations of the system.

\subsection{\label{sec:lev8}The low loading limit.}
In this case we can assume that there is never more than one particle in
the system. The following rate equations then hold.
\begin{equation}
\begin{split}
  {d\expec{\rm A}\over dt}
  &={2W_{\rm ads}\over S}\expec{*}
   -{2W_{\rm des}\over S}\expec{\rm A}\cr
  & -{W_{\rm rx}\over 2N_{\rm inert}+1}\expec{\rm A}\cr
  {d\expec{\rm B}\over dt}
  &=-{2W_{\rm des}\over S}\expec{\rm A}
   +{W_{\rm rx}\over 2N_{\rm inert}+1}\expec{\rm A}\cr
\end{split}
\end{equation}
Here $\expec{\rm X}$ is the probability that there is an X at an
arbitrary site. For steady state we get
\begin{equation}
\begin{split}
  \expec{\rm A}
  &={2W_{\rm ads}W_{\rm des}
  \over (W_{\rm ads}+W_{\rm des})
  (2W_{\rm des}+N_{\rm prot}W_{\rm rx})}\cr
  \expec{\rm B}
  &={N_{\rm prot}W_{\rm ads}W_{\rm rx}
  \over (W_{\rm ads}+W_{\rm des})
  (2W_{\rm des}+N_{\rm prot}W_{\rm rx})}\cr
\end{split}
\end{equation}
From this we immediately get the turnover frequency
\begin{equation}
  \omega_{\rm TOF}={2W_{\rm ads}W_{\rm des}W_{\rm rx}
  \over (W_{\rm ads}+W_{\rm des})
  (2W_{\rm des}+N_{\rm prot}W_{\rm rx})}.
\end{equation}
We see that in this limit the turnover frequency does not depend on the
system size.

 Comparing the number of $\rm B$'s produced from the analytical results
with the DMC results for the case when $W_{\rm ads}=0.0033$, $W_{\rm
des}=0.9967$, $W_{\rm rx}=0.1$ and different distributions of the reactive
sites, we remark that we get similar results.



%
%
%
\subsection{\label{sec:lev9} Fast and slow ${\rm A}\to{\rm B}$ reaction.}
If the reaction is fast, and there are not too many particles in the
system, then all particles in the system are Bs. This means
\begin{equation}
  \expec{\rm B}={W_{\rm ads}\over W_{\rm ads}+W_{\rm des}},
\end{equation}
and
\begin{equation}
  \omega_{\rm TOF}
  ={1\over N_{\rm prot}}
  {2W_{\rm ads}W_{\rm des}\over W_{\rm ads}+W_{\rm des}}.
\end{equation}
The restriction of not too many particles is necessary, because
particles should all always be able to reach a protonic site. This means
\begin{equation}
  {W_{\rm ads}\over W_{\rm ads}+W_{\rm des}}
  \ll{N_{\rm inert}\over S}
\end{equation}
must hold.
 This relation depends on the distribution of the reactive sites. When
reaction is fast this means that it depends on the distance from the margins
to the first protonic site.

 Comparing the site occupancy with $\rm B$'s(20)  with results from the DMC
simulations, we find that for the reaction rate constants ($W_{\rm ads}$=0.03333,
$W_{\rm des}$=0.96667, $W_{\rm rx}$=10 and all the sites reactive, the
results are similar.



If the reaction is slow, then there are only occasionally $\rm B$'s in the
system. This means
\begin{equation}
  \expec{\rm A}={W_{\rm ads}\over W_{\rm ads}+W_{\rm des}}.
\end{equation}
All particles in the system will be renewed between two subsequent
formations of a B. Therefore
\begin{equation}
  \omega_{\rm TOF}
  ={W_{\rm ads}W_{\rm rx}\over W_{\rm ads}+W_{\rm des}}.
\end{equation}



 Comparing the site occupancy with $\rm A$'s(23) with results from the DMC
simulations, we find that for the reaction rates constants $W_{\rm ads}$=0.03333, $W_{\rm des}$=0.96667,
$W_{\rm rx}$=0.001 and all the sites reactive, the results are similar.


\section { Simulation Results and Discussion }

\subsection {\label{sec:lev10} All sites reactive}	

\begin {figure}
\centering
\subfigure {\epsfig {figure=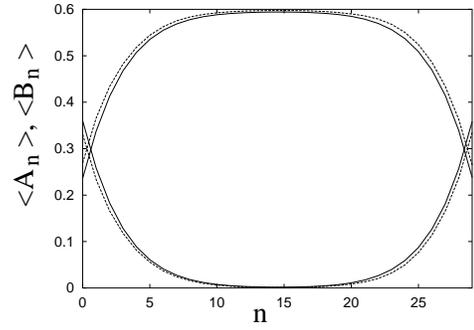, width=6.5cm} }
\caption {
           Profile occupancies for a system of length $S=30$, 
           $W_{\rm ads}$=0.6, $W_{\rm des}$=0.4 and $W_{\rm rx}$=0.1. The continous
           lines are the profile occupancies for $\rm A$ (the lower) and
	   $\rm B$ (the higher)
           using DMC for infinitely-fast diffusion. The dotted lines are the
           profile occupancies for $\rm A$(the lower) and $\rm B$ (the higher) using 
           DMC with a regular high rate for diffusion ($W_{\rm diff}=1600$).
         }
\label{carlos_infin_prof}
\end{figure}

 We remark that DMC methods with regular high rates for diffusion tend to
give similar results as DMC method for infinitely-fast diffusion described in
section ~\ref{sec:lev5}. The results of this comparisons are in figure
~\ref{carlos_infin_prof}. We conclude that the DMC method for infinitely-fast 
diffusion we have defined is a correct method to simulate the behavior of
the system in the limit $W_{\rm diff} \to \infty$.

 In~\cite{silvia} we have seen that for the case when all the sites are
reactive, the site occupancies of the system obtained from DMC simulations
show that the system is not homogeneous even for very fast diffusion rates. We find the same effect also 
using DMC for infinitely-fast diffusion, for different loadings and for
different reaction rate constants. 

 We study also the dependence of 
the occupancy profiles on the reaction rate constant at 
different loadings of the system.
 
\begin {figure}
\centering
\subfigure {\epsfig {figure=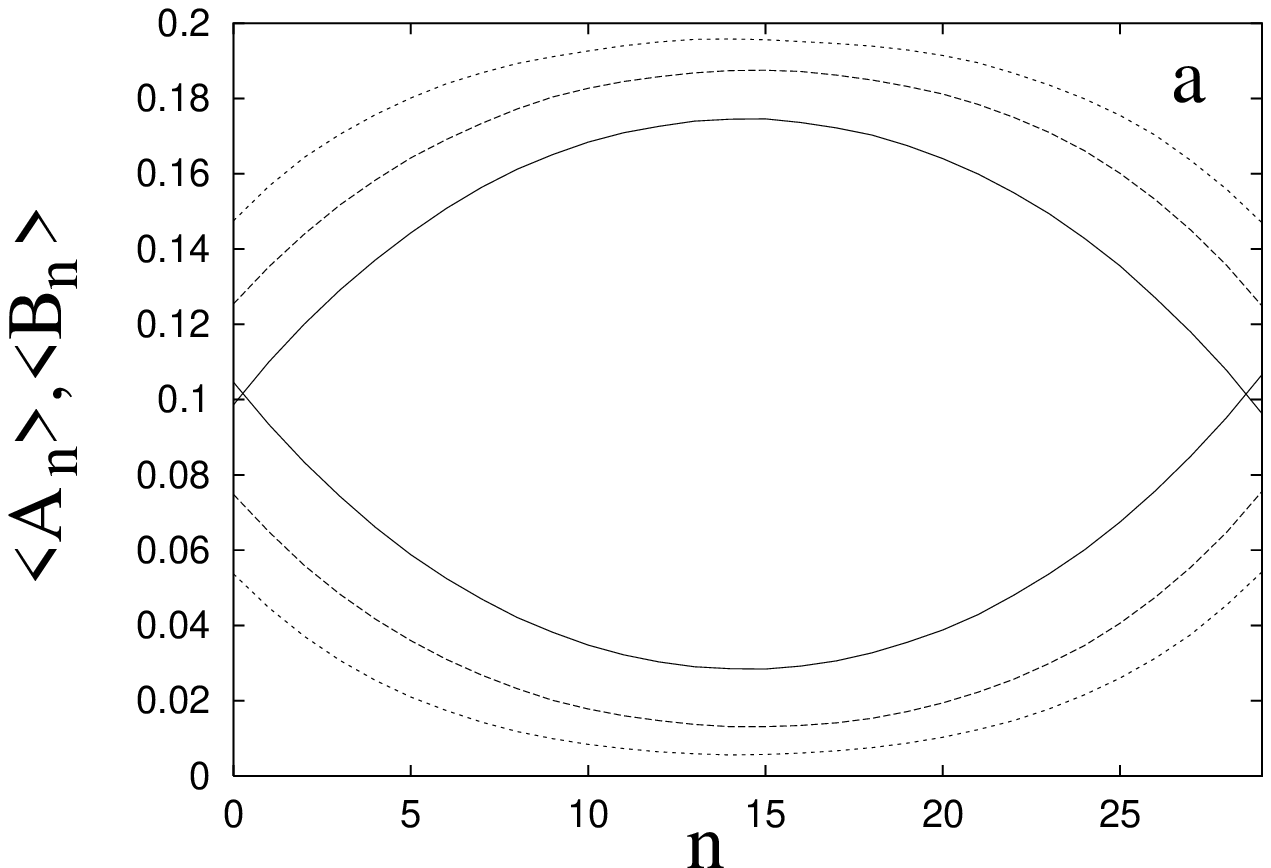, width=5.5cm} }
\subfigure {\epsfig {figure=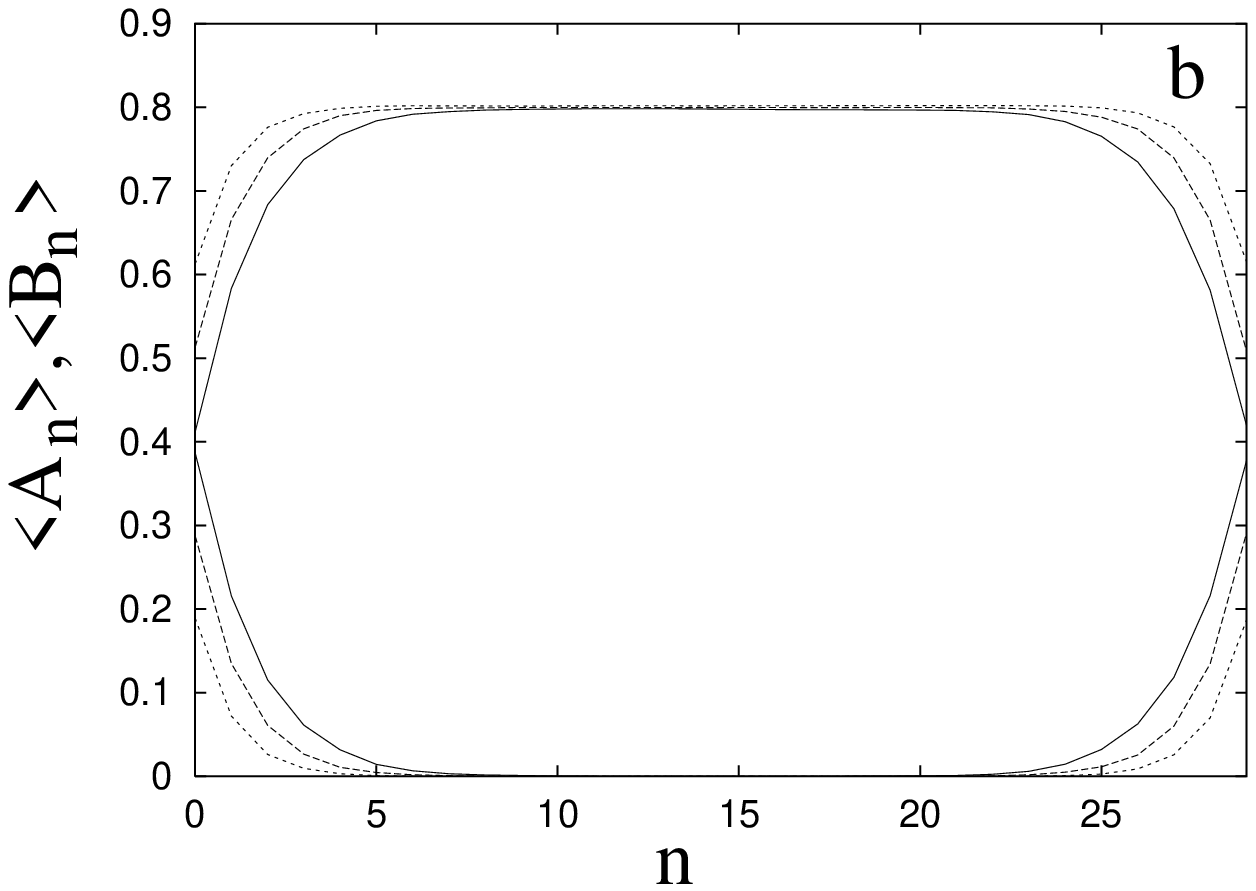, width=5.5cm} }
\caption {
          a)Dynamic Monte Carlo results for site occupancy for the case of 
           infinitely-fast diffusion and low loading
           ($W_{\rm ads}=0.2$, $W_{\rm des}=0.8$) of a system of length $S=30$.
            The continuous line is for the site occupancy for $W_{\rm rx}=0.1$,
            the first dotted line near the continous line is for $W_{\rm rx}=0.2$,
            and the second is for $W_{\rm rx}=0.4$.
          b)Dynamic Monte Carlo results for site occupancy for the case of 
           infinitely-fast diffusion and high loading
           ($W_{\rm ads}=0.8$, $W_{\rm des}=0.2$) of a system of length $S=30$.
            The continuous line is for the site occupancy for $W_{\rm rx}=0.1$,
            the first dotted line near the continous line is for $W_{\rm rx}=0.2$,
            and the second is for $W_{\rm rx}=0.4$.
         }
\label{carlos_prof_infin_rx}
\end{figure}

  The simulation results in figure~\ref{carlos_prof_infin_rx} 
show how the shape of the profiles changes with reaction rate $W_{\rm rx}$ for 
high and low loading of the pipe, when all the sites are reactive. We find that for high loadings, as an
effect of the blocking, the middle sites have the same probability to be
occupied for both fast and slow reactive system. Only the occupancy of
marginal sites is influenced by the reactivity, such that in fast reactive
systems we have a higher probability to have a $\rm B$ near the marginal
sites, and, in consequence, a better productivity. 
 For slow reactive systems, the occupancy profiles are scaled with reaction
rate constant. We notice that the productivity is growing 
increasing reaction rate in case of low loading systems almost as fast as in
the case of the high loading systems because of the diminished effect of the
blocking in the pipe. Comparing for instance the rate o growth for $\rm B$
production when reaction rate constant is increasing from $W_{rx}=0.1$ to
$W_{rx}=0.4$ ($B_{prod}^{W_{rx}=0.4}-B_{prod}^{W_{rx}=0.1})/S$), in case of 
low loading ($\theta=0.2$) and high loading ($\theta=0.8$), we find
almost the same rate of growth in both the cases, and this is 0.5.

\begin {figure}
\centering
\subfigure {\epsfig {figure=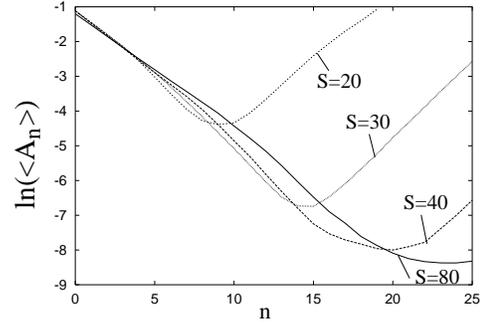, width=6.5cm} }
\caption { 
  The logarithm of the DMC(Random Selection Method) profile occupancies
  ($\langle {\rm A}_n\rangle$) for $W_{\rm ads}=0.6$, $W_{\rm des}=0.4$,
   $W_{\rm rx}=0.1$, and when all
  the sites are reactive, for various system sizes $S$.
}
\label{COMP_LN_carlos.eps}
\end{figure}

In figure~\ref{COMP_LN_carlos.eps}  we have the logarithmic shape of the
occupancy profiles for $\rm A$ and $\rm B$. These profiles 
show that smaller systems are less reactive as less $\rm A$s are inside the 
pipe. This explains the faster decrease of $\langle {\rm A}_n \rangle$ for small 
systems in figure~\ref{COMP_LN_carlos.eps}. For infinitely-long systems we 
expect to have straight lines corresponding to an exponential decrease of 
$\langle {\rm A}_n \rangle$.
 
\subsection {\label{sec:lev11} Only some of the sites reactive }

\begin {figure*}
\centering
\subfigure {\epsfig {figure=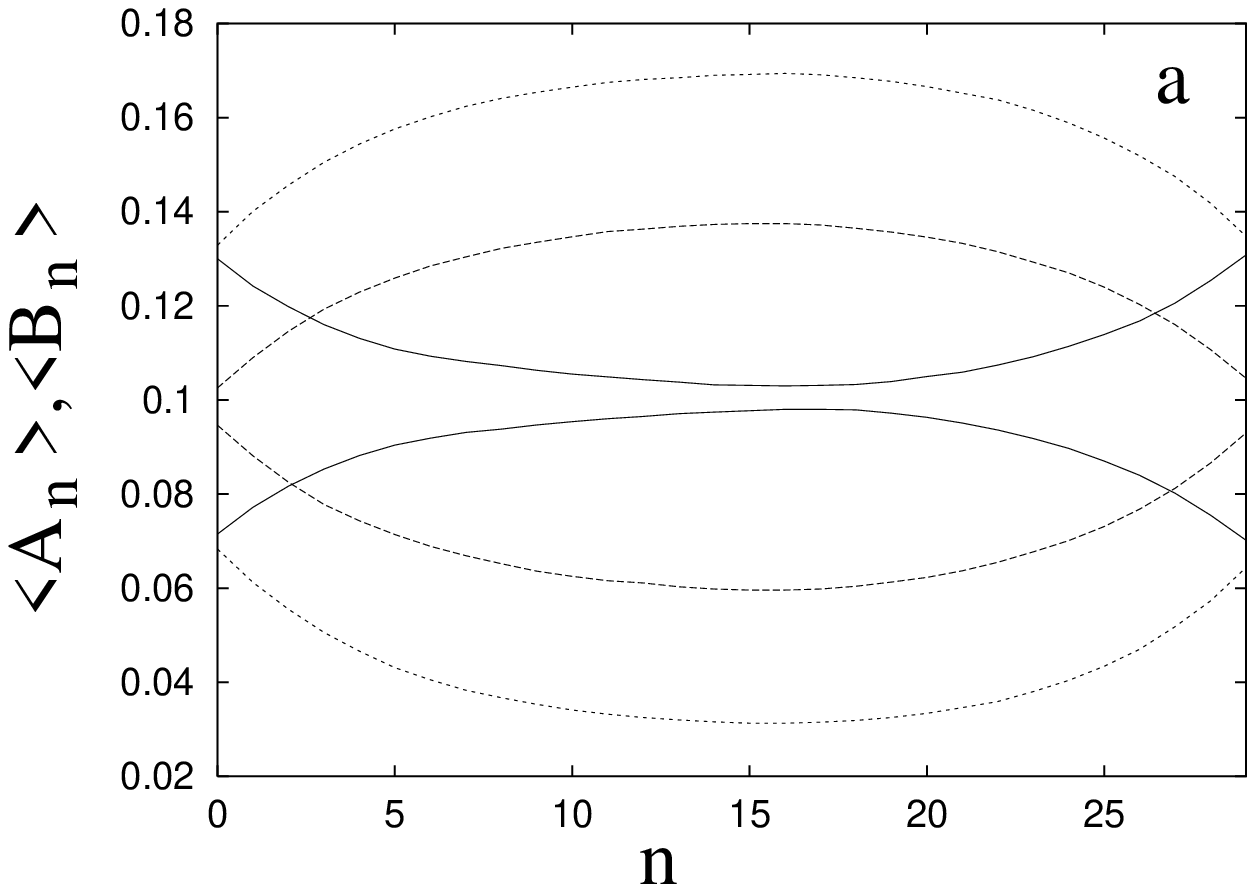, width=4.5cm} }
\subfigure {\epsfig {figure=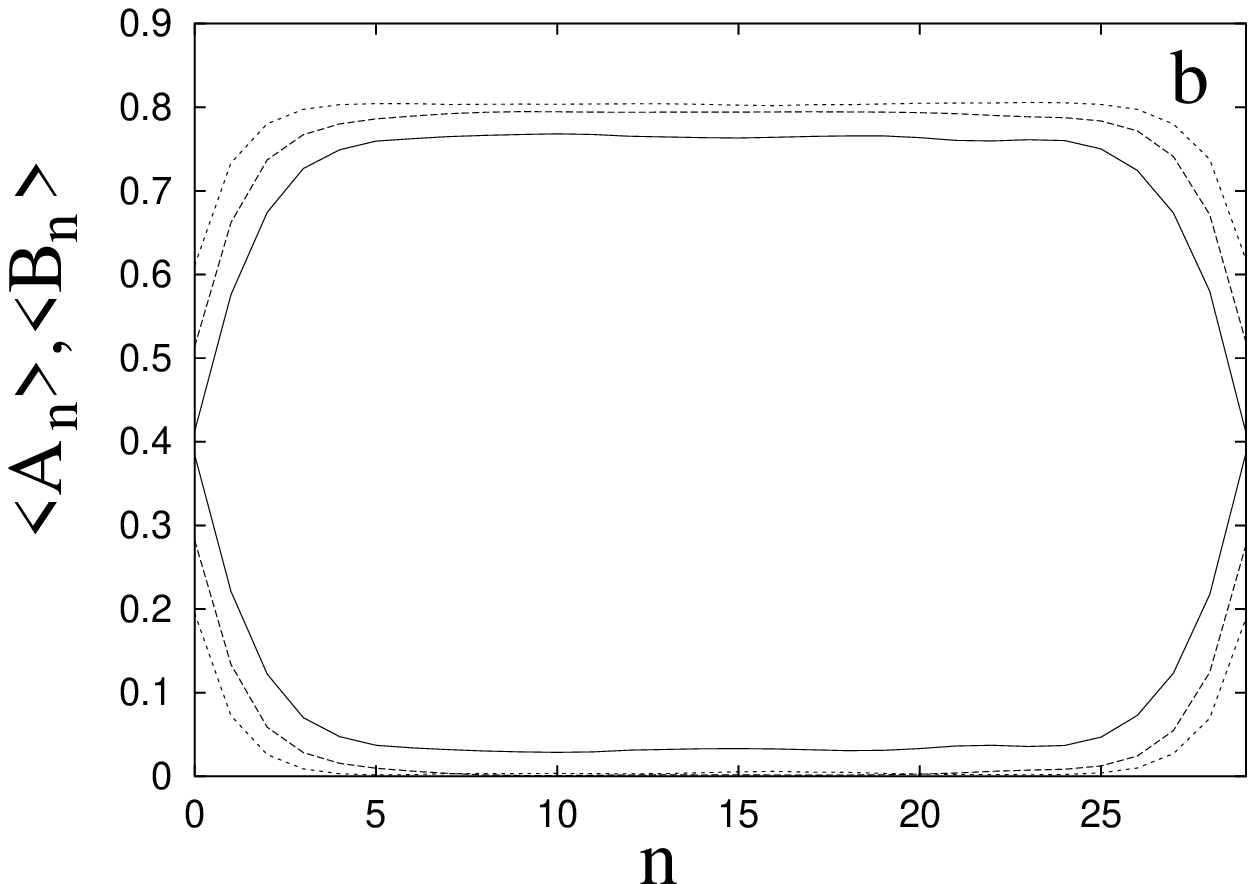, width=4.5cm} }
\subfigure {\epsfig {figure=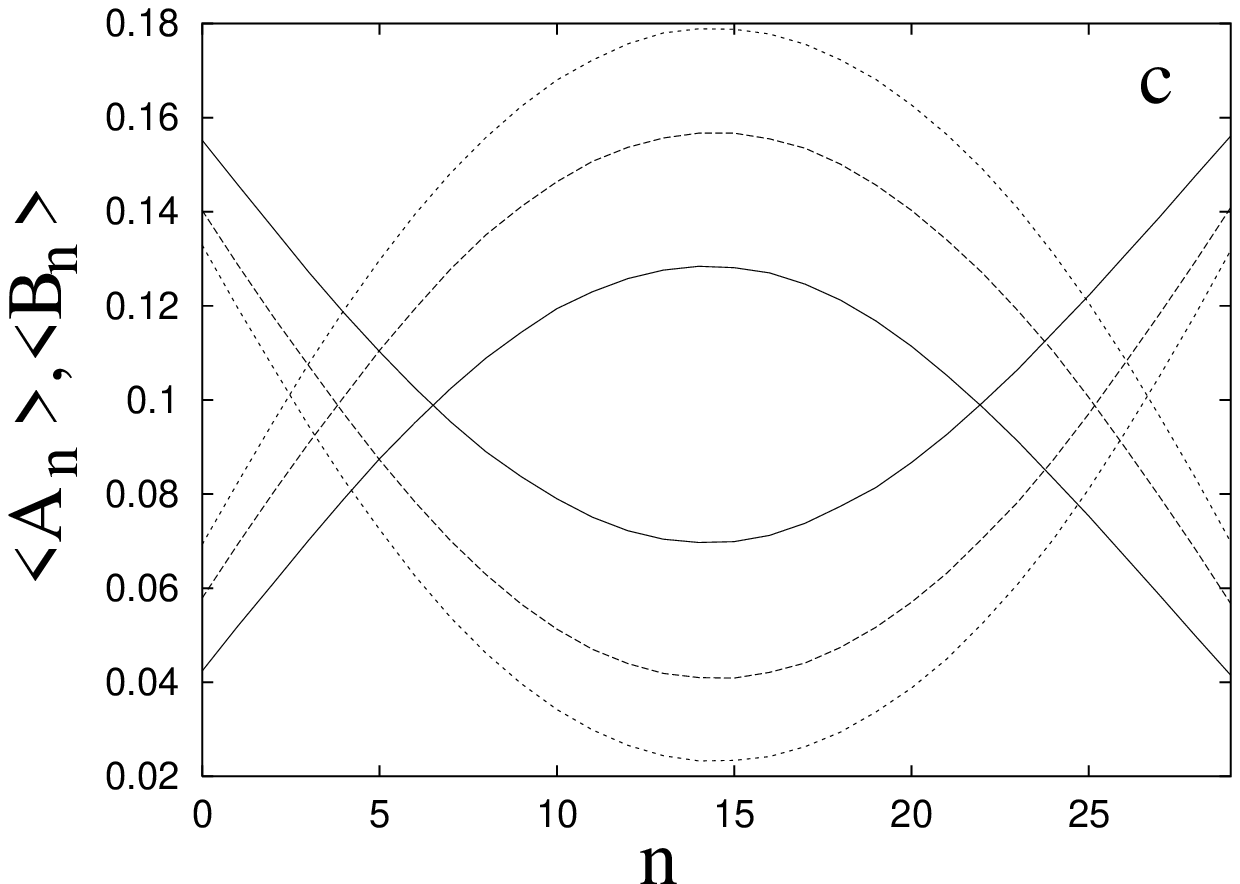, width=4.5cm} }
\subfigure {\epsfig {figure=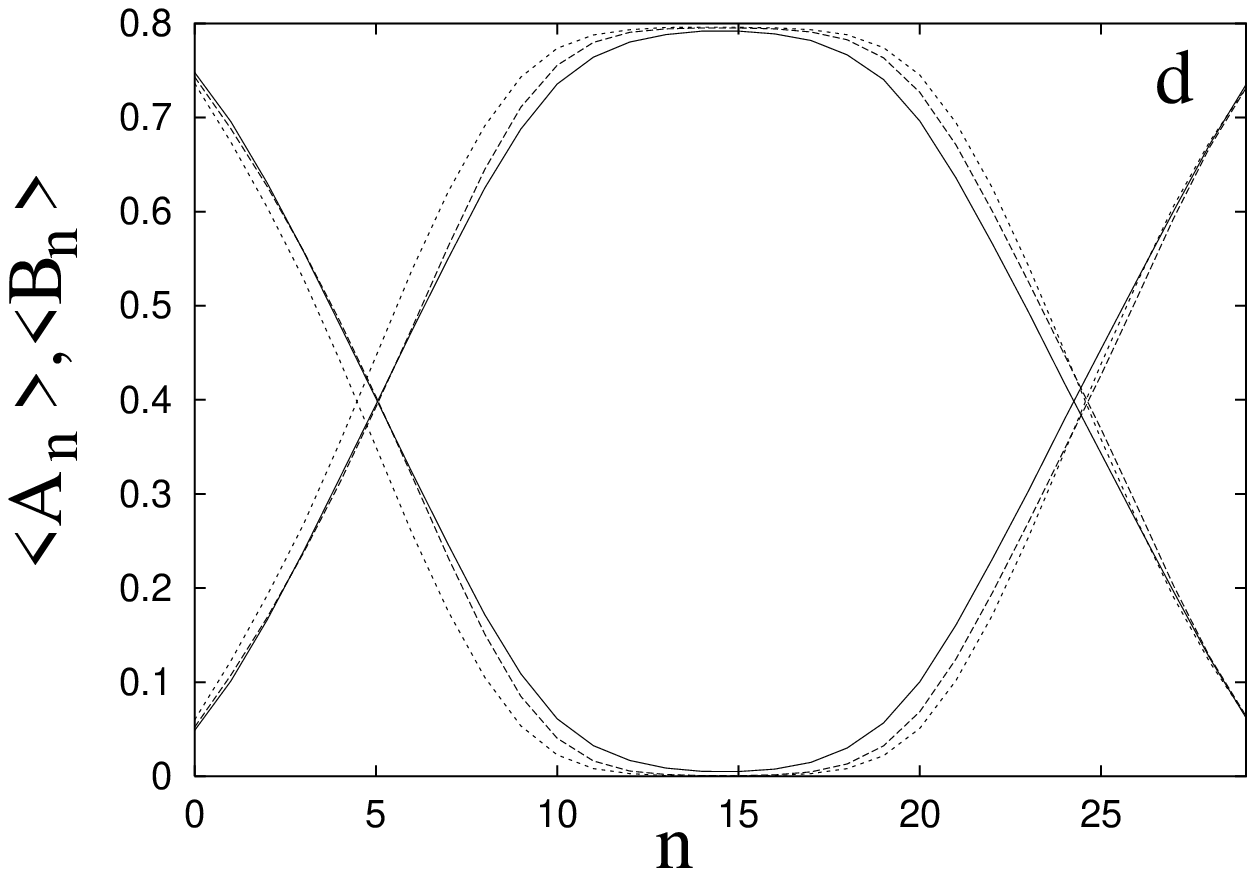, width=4.5cm} }
\caption { Dynamic Monte Carlo results for site occupancy for the case of 
           infinitely-fast diffusion.The first two
           figures(a,b) are for the case of 5 marginal sites reactive at each end and the
           last two(c,d) for the case of 10 middle sites reactive. The first figure is
           for the case of low loading ($W_{\rm ads}=0.2$, $W_{\rm des}=0.8$),
           and the second for high loading ($W_{\rm ads}=0.8$, $W_{\rm des}=0.2$) 
           at different reaction rates ($W_{\rm rx}=0.1, 0.2, 0.4$).
           The third and the fourth figures are for the same parameters as
           the first and the second, but for the case of 10 middle sites
           reactive.
         }
\label{carlos_prof_infin_rx_marg}
\end{figure*}

  For all the sites reactive, in paper ~\cite{silvia}, we have shown that even
 when $W_{diff} \to \infty$, DMC results with regular high diffusion rates indicate that
the system doesn't become homogeneous.
 Using DMC for infinitely-fast diffusion we find that also for different
distribution of the reactive sites we find a non-homogeneous distribution of
the particles in the system.
 In figure~\ref{carlos_prof_infin_rx_marg} we can see that for marginal
sites reactive and for middle sites reactive as well, we have single-file
effects also for infinitely-fast diffusion for different rates of reactions.

 We notice that the profiles in case of high loadings are very slowly
dependent on conversion, both for middle and for marginal sites reactive.
 Comparing with simulation results for occupancy profiles in ~\cite{silvia}, 
where middle sites and marginal sites are reactive, and for regular diffusion rate
constants in the domain $(2 \ldots 10)$, we notice that the profiles are similar. 
 We can conclude that as the effect of infinitely-fast diffusion is absent for
Single-File Systems, the diffusion is not so important in these systems.

 For low loading, when marginal sites
are reactive, the occupancy profiles are scaled with $W_{\rm rx}$.
 We notice that conversion in figure~\ref{carlos_prof_infin_rx_marg} is the
rate determining step. In this
case, the middle sites doesn't have the same occupancy for different
reaction rate constants like in the case of all the
sites reactive but are strongly dependent on $W_{rx}$.

We notice also that the productivity in case marginal
sites are reactive is growing faster increasing the reaction rate constants at
low loadings than at high loadings. Comparing, for instance, the rate o growth 
for $\rm B$ production $B_{prod}$ when reaction rate constant is increasing from $W_{rx}=0.1$ to
$W_{rx}=0.4$, in case of low loading ($\theta=0.2$) and 
high loading ($\theta=0.8$), we find that the rate of growth of $\rm B$ 
productivity at low loadings ($0.3$) is larger than at high loadings 
($0.25$).
 
 When we have middle sites reactive we have higher probability to find 
$\rm A$'s near marginal sites. The productivity is smaller than in all the
other cases. The profiles are again scaled with $W_{\rm rx}$
for low loadings. For high loadings and middle sites reactive, the profiles
for different conversion rates are almost similar, so the productivity can
only be increased in this case only increasing the number of reactive sites.

\section {Summary}

  We have used DMC methods and analytical techniques to analyze Single-File
Systems for which diffusion is infinitely-fast.
  We simplified the ME removing fast diffusion and we have presented a DMC
algorithm for infinitely-fast diffusion that simulate this ME. We show that
DMC with regular high rates gives the same results as DMC for
infinitely-fast diffusion.
 The fundamental assumption  considered for infinitely-fast diffusion in the
 analytical results is that all configurations related by diffusion
have the same probability.

 In the limiting cases such as low loading limit and
slow and fast conversion, we are able to derive expressions for the 
$\rm B$ productivity. We notice that the number of $\rm B$s produced per
unit time in these cases doesn't depend on the system size. Comparisons
between analytical and DMC results reveal similar results for the
productivity.




  DMC results show that when all the sites are reactive and when only some of the sites
are reactive, diffusion has no influence on the single-file
properties of the system. Different results for the dependencies of the occupancy profiles
and productivity on the reaction rate constant and different distributions
are categorized. The occupancy profiles show that smaller systems are less
reactive as less $\rm A$s are inside the pipe.



\bibliographystyle{./apsrev}
\nocite{*}

\end {document}